\documentclass[12pt]{article}

\usepackage{latexsym}
\usepackage{amssymb,cite}
\usepackage{amsmath,epsfig}
\usepackage{rotating}
\usepackage{graphics,color}
\usepackage{a4}

\def\be#1\ee{\begin{equation}#1\end{equation}}

\def\be#1\ee{\begin{equation}#1\end{equation}}
\def\bea#1\eea{\begin{align}#1\end{align}}

\begin{document}

\title{\bf\large 
\vspace*{-2cm}
\begin{flushright}
%{\normalsize\normalfont MPP-2011-3}
\vspace*{1cm}
\end{flushright}
On the Higgs-Confinement Complementarity \\
}

\author{
Erhard Seiler$^a$\footnote{email: ehs@mpp.mpg.de}  \\
\mbox{} \\
\mbox{} \\
 $^a${\em\normalsize Max-Planck-Institut f\"ur Physik 
(Werner-Heisenberg-Institut)} \\
 {\em\normalsize M{\"u}nchen, Germany} \\
}

%\date{June 1st, 2015}
\date{\today}

\maketitle

\begin{abstract} \noindent It has been noticed long ago that in Higgs 
models with `complete symmetry breaking' one can move from the confinement 
to the Higgs regime without crossing a phase boundary, a fact sometimes 
called referred to as `complementarity'. In a recent paper some doubt was 
raised about the correctness of the mathematics underlying this fact and 
it was claimed that the supposed `flaw' would resolve the `paradox' seen 
in this complementarity. Here we briefly revisit the facts both from a 
mathematical and a physical point of view and point out that (a) there is 
no paradox and (b) there is no flaw in the mathematical reasoning.

\end{abstract}

\maketitle

\newpage

\section{Introduction}

After K.~Wilson \cite{Wilson:1974sk} created lattice theory and thereby 
found a well-defined nonperturbative formulation of gauge theories (albeit 
with an ultraviolet cutoff) that did not require gauge fixing, it became 
possible to address questions not answerable in perturbation theory. The 
first and most important such question, the question of confinement of 
quarks, was answered by Wilson himself (with later mathematical refinement 
in \cite{Osterwalder:1977cv}) for strong enough bare coupling.

Very soon the interest turned to the Higgs mechanism, which in 
perturbation theory with gauge fixing is conventionally described as 
spontaneous symmetry breaking (SSB). In \cite{Osterwalder:1977cv} it was 
shown using a cluster expansion that a Yang-Mills-Higgs model having a 
property conventionally denoted as `complete breakdown of symmery' has a 
mass gap at arbitrarily weak gauge coupling, provided the Higgs potential 
is sufficiently strong. This result was analyzed in \cite{Fradkin:1978dv} 
and physically interpreted as the absence of a phase boundary between 
`confinement' and `Higgs' regions, a property that became known as 
`complementarity'.

Lately is has been been claimed by Grady \cite{Grady:2015dsa} that the 
proof of the relevant theorem in \cite{Osterwalder:1977cv} contains a 
flaw, that the theorem is invalid and thereby a `paradox' is solved. Here 
we will first remind the reader why there is no paradox; to set the record 
straight we then briefly revisit the arguments entering the proof and show 
why there is no flaw.

\section{There is no paradox} 

G.~'t Hooft \cite{'tHooft:1979bi} in 1979 explained confinement-Higgs 
complementarity in physical terms. He showed that the Higgs mechanism may 
just as well be interpreted as confinement: bare Yang-Mills quanta and 
bare Higgs quanta may be combined into gauge invariant, permanently bound 
compounds forming the gauge invariant physical `W' and Higgs bosons, just 
as quarks and gluons are bound into hadrons.  This is gratifying, because 
from Elitzur's theorem \cite{Elitzur:1975im} we know that gauge invariance 
cannot be broken spontaneously if no gauge fixing term is present. This 
fact is in some sense obvious, because the local invariance means that 
there is no preference for aligning even nearby degrees of freedom 
(i.~e.~there is not even short range order).

A little later Fr\"ohlich, Morchio, and Strocchi \cite{Frohlich:1981yi} 
set out to formulate perturbation theory for such models in terms of gauge 
invariant composite fields without the usual assumption of SSB via a 
nonvanishing expectation value of the Higgs field. The procedure outlined 
there is not necessarily useful for computational purposes, but it is 
important to recognize that it is possible in principle.  Fr\"ohlich et al 
also showed in that paper that SSB is absent even in the presence of 
certain gauge fixes.

Another important, unfortunately much too little recognized result is due 
to Kennedy and King \cite{Kennedy:1985yn}. They considered abelian lattice 
gauge theories in 4 dimensions with a covariant gauge fixing term, as 
employed usually in continuum perturbation theory. There finding was: SSB 
occurs {\it only} in the {\it Landau gauge}, i.e. the gauge with minimal 
infrared fluctuations. In all other covariant gauges infrared fluctuations 
destroy SSB, much like infrared fluctuations destroy long range ordering 
in 2D ferromagnets (a fact known as the Mermin-Wagner theorem 
\cite{Mermin:1966fe}).

Provocately one might say that SSB in Higgs models, if it occurs at all, 
is a gauge fixing artefact. 

\section{There is no flaw in the proof of analyticity} 

Here want to briefly review the structure of the proof of analyticity 
given in \cite{Osterwalder:1977cv}, in order to clear up some 
misconceptions. It proceeds by a convergent cluster expansion in the form 
developed by Glimm, Jaffe and Spencer \cite{gjs} in the context of 
Constructive Quantum Field Theory. This expansion expresses expectation 
values first as a sum of finite volume quantities which are clearly 
analytic in the parameters; then it is shown that the expansion converges 
uniformly in the volume in a certain complex domain of parameters. This by 
itself is, however, not sufficient to prove analyticity of the infinite 
volume limit, as correctly remarked by Grady (it is also not sufficient to 
establish the existence of that limit). Grady goes on to suggest that the 
limits of the individual terms might be nonanalytic.

But in \cite{Osterwalder:1977cv} it is remarked that existence of the 
infinite volume limit and its analyticity follow by standard applications 
of the cluster expansion, for which the authors refer to \cite{gjs}. 
Obviously the authors of \cite{Osterwalder:1977cv} did not want to repeat 
the arguments given in \cite{gjs}, as they were considered `standard' in 
their community. So let me repeat the structure of the argument for 
analyticity of the infinite volume limit:

(1) The cluster expansion converges uniformly (in the volume) in a certain 
complex domain in the parameters of the model (the conditions necessary 
for lattice gauge theories are given in OS).

(2) From this follows uniform (in the volume) exponential clustering for 
pairs of observables.

(3) This clustering in turn implies convergence to the thermodynamic limit
because boundary effects decouple exponentially.

(4) Vitali's convergence theorem for holomorphic functions then implies
analyticity of the limit in the domain specified before.

It should also be said that there are other, possibly more transparent 
ways of arriving at the results; in particular so-called polymer 
expansions described in \cite{Seiler:1982pw} and later in the textbook 
\cite{Glimm:1987ng} produce a somewhat different series; the main 
advantage of this approach is the fact that every term in the series 
reaches its thermodynamic limit already at a certain finite volume, so
analyticity of each term is obvious, because it is equal to a finite 
volume quantity.  

\section{Conclusions}

On the lattice there is, at least in the case of `complete symmetry 
breakdown', no phase boundary completely separating the `Higgs' and 
`Confinement' regimes (just as in the liquid-gas case); the old proofs are 
valid. This is, however, not a paradox; the phenomenological explanation 
provided by 'Hooft makes the fact intuitively plausible.

\end{document}